\documentclass[11pt]{article}

\usepackage[T1,T2A]{fontenc}
\usepackage[cp1251]{inputenc}
\usepackage{textcomp}
\usepackage[centertags]{amsmath}
\usepackage{amsfonts}
\usepackage{amssymb}
\usepackage[pdftex]{hyperref}
\usepackage{graphicx}
\usepackage{graphbox}
\usepackage[numbers,sort&compress]{natbib}

\usepackage{paperinitial}

\paperinitialization{15mm}{15mm}{15mm}{15mm}{2pt}{10pt}

\DeclareMathOperator{\re}{Re}

\newcommand{\vf}{\varphi}
\newcommand{\vk}{\varkappa}
\newcommand{\s}{\sigma}

\newcommand{\al}{\alpha}
\newcommand{\be}{\beta}
\newcommand{\ga}{\gamma}
\newcommand{\Ga}{\Gamma}
\newcommand{\de}{\delta}
\newcommand{\De}{\Delta}

\newcommand{\la}{\lambda}
\newcommand{\La}{\Lambda}

\newcommand{\spx}{\mathbf{x}}

\begin{document}
\allowdisplaybreaks[4]
\frenchspacing
\setlength{\unitlength}{1pt}

\title{{\Large\textbf{Probability of radiation\\ of twisted photons by axially symmetric bunches of particles}}}

\date{}

\author{O.V. Bogdanov${}^{1),2)}$\thanks{E-mail: \texttt{bov@tpu.ru}},\; P.O. Kazinski${}^{1)}$\thanks{E-mail: \texttt{kpo@phys.tsu.ru}}\\[0.5em]
{\normalsize ${}^{1)}$ Physics Faculty, Tomsk State University, Tomsk 634050, Russia}\\
{\normalsize ${}^{2)}$ Tomsk Polytechnic University, Tomsk 634050, Russia}}

\maketitle

\begin{abstract}

The effect of a finite width of a particle bunch on radiation of twisted photons is studied. The general formulas connecting the radiation probability distribution of twisted photons produced by bunches of identical particles with the radiation probability distribution of twisted photons generated by one particle are obtained for axially symmetric bunches. The bunch is called axially symmetric if it is axially symmetric with respect to the detector axis at some instant of time and all the particles in the bunch move along parallel trajectories. The general sum rules for the probability of radiation of twisted photons by axially symmetric bunches are established. In particular, we prove that the projection of the average total angular momentum of radiated twisted photons per particle in the bunch does not depend on the radial profile of the bunch. The uniform, Gaussian, and exponential radial bunch profiles are considered in detail. The radiation of axially symmetric bunches in ordinary and crystalline undulators is investigated. The selection rules for radiation of twisted photons by one particle in undulators are violated when the finite width of the particle bunch is taken into account. We find the condition when this violation is marginal. The form of the radiation probability distribution of twisted photons becomes universal for wide incoherent axially symmetric particle bunches. We completely describe these universal distributions.

\end{abstract}

\section{Introduction}

The twisted photons are the states of the electromagnetic field that possess definite the energy, the projections of the momentum and the total angular momentum onto the detector axis, and the helicity \cite{GottfYan,JaurHac,BiaBirBiaBir,JenSerprl,JenSerepj,Ivanov11}. These states constitute a complete set, and any free electromagnetic field can be represented as a collection of twisted photons. It is important that such decomposition is not only a mathematical property, but it can be performed by suitably designed detectors and observed experimentally \cite{LPBFAC,BLCBP,SSDFGCY,LavCourPad,RGMMSCFR}. The wave packets of twisted photons sufficiently narrow in quantum numbers are widely used now in science and technology (see for reviews \cite{PadgOAM25,AndBabAML,TorTorTw,AndrewsSLIA}).

One of the bright and sufficiently pure sources of twisted photons are undulators and undulator-like devices, for example, FELs and laser waves. The generation of narrow wave packets of twisted photons by such devices was predicted theoretically \cite{SasMcNu,HMRR,HeMaRo,AfanMikh,BordKN,SSFS,Sherwin,KatohPRL,TaHaKa,Rubic17,CLHK,JenSerprl,JenSerepj} and confirmed experimentally \cite{HKDXMHR,BHKMSS,Rubic17,KatohSRexp}. Recently, we have found a general formula for the probability of radiation of twisted photons by classical currents \cite{BKL2}. This formula allowed us to describe completely the radiation produced by one point charged particle in undulators and wigglers, but the effects caused by a finite size of a particle bunch were completely neglected. In contrast to radiation of plane-wave photons, the radiation probability of twisted photons is not invariant under translations of a source that are perpendicular to the detector axis \cite{BKL2}. Therefore, a finiteness of the particle bunch results in nontrivial modification of the radiation probability of twisted photons even in the case of incoherent radiation. In the present paper, we investigate the effect of finiteness of the transverse bunch size on the form of the probability of radiation produced by axially symmetric bunches (see the precise definition below). We establish that, in general, the finite transverse size of a particle bunch leads to broadening of the radiation probability distribution, over the projection $m$ of the photon total angular momentum onto the detector axis, in comparison with the radiation created by one point particle.

In Sec. \ref{Gener_Form}, we start with the general formulas for the probability of radiation of twisted photons produced by a bunch of identical particles that is axially symmetric with respect to the detector axis at some instant of time and all the trajectories of particles in the bunch are obtained from one trajectory by a parallel transport. For brevity, we call such a bunch as axially symmetric. Such round bunches possess a high bunch-by-bunch luminosity and are created, for example, in the electron-positron collider VEPP-2000, Novosibirsk \cite{RPPPDG2018}. The assumption that the bunch is axially symmetric allows us to obtain simple formulas relating the radiation probability for a bunch to the radiation probability for one point particle. These formulas involve the interference factors depending on the radial distribution of particles in the bunch. Several exact sum rules for the probability distributions and interference factors are found. In particular, it is shown that the radiation produced by an axially symmetric bunch and the radiation produced by one particle possess the same projections of the total angular momentum onto the detector axis per photon. This property holds for both coherent and incoherent radiations of a bunch. As for incoherent radiation, this sum rule is valid for bunches with zero dipole moment with respect to the center of the bunch and not only for axially symmetric bunches, provided the particles move along parallel trajectories. Then we derive a simple general condition when the one-particle answer for radiation probability can be used to describe radiation produced by a bunch. As an example, we obtain the interference factors for bunches with simple radial profiles: uniform, Gaussian, and exponential. Then we consider the generalization of formulas to the case when the symmetry axis of the bunch is tilted from the detector axis by a small angle. In that case, the formula for incoherent radiation remains intact, while the formula for coherent radiation is modified and the new interference factor arises. We find the explicit expression for this factor for bunches with simple radial profiles mentioned above. In conclusion of Sec. \ref{Gener_Form}, the transition from the bunch of particles to a continuous flow is discussed.

In Sec. \ref{Und_Rad}, we apply the general formulas to describe the radiation of twisted photons by axially symmetric bunches of charged particles in undulators and wigglers. We show that the selection rules fulfilled for the forward radiation generated by one particle in undulators \cite{Rubic17,KatohPRL,TaHaKa,SasMcNu,BKL2} become violated when the radiation produced by a bunch of such particles is considered. However, if the spreading of radiation probability distribution over $m$ is small, this violation is marginal. We find the condition when broadening of the distribution over $m$ is negligible. As the second example, we consider the radiation of twisted photons by crystalline undulators \cite{KKSG,KKSGb}. The possibility of creation of radiation with large angular momentum by channeling of charged particles was pointed out in \cite{ABKT} (see also \cite{EpJaZo}). Here we study a relatively soft radiation generated by particles moving in a specially designed crystal. The spreading of radiation probability distribution over $m$ is very large in this case and the shape of this distribution becomes universal. Thus, in considering the undulator radiation of twisted photons, we examine their radiation in a wide range of energies from the THz domain to the hard X-rays. In conclusion section, we summarize the results.

We use the notation and conventions adopted in \cite{BKL2}. In particular, $\hbar=c=1$ and $e^2=4\pi\al$, where $\al\approx1/137$ is the fine structure constant.

\section{General formulas}\label{Gener_Form}

\paragraph{Translations of the source.}

Let us find the change of the radiation amplitude under translations of the source by the vector $\mathbf{a}=(a_1,a_2,a_3)$. Suppose that the translations are realized in the Fock state space by a unitary operator $\hat{V}(\mathbf{a})$. Then
\begin{equation}
    \hat{V}(\mathbf{a})\hat{A}_i(\spx)\hat{V}^{-1}(\mathbf{a})=\hat{A}_i(\spx+\mathbf{a}),
\end{equation}
where $\hat{A}_i(\spx)$ is the operator of the electromagnetic potential in the Coulomb gauge. Substituting the expansion of the operator $\hat{A}_i$ in terms of the mode functions [(18), \cite{BKL2}] into this expression, we obtain
\begin{equation}
    \hat{V}(\mathbf{a})\hat{c}_\al\hat{V}^{-1}(\mathbf{a})=\La_{\al\be}(\mathbf{a})\hat{c}_\be,\qquad\psi_{\al i}(\spx+\mathbf{a})=\psi_{\be i}(\spx)\La_{\be\al}(\mathbf{a}),
\end{equation}
where $\psi_{\al i}(\spx)$ are the mode functions of a twisted photon [(17), \cite{BKL2}]. The coefficients $\La_{\al\be}(\mathbf{a})$ can easily be found with the aid of the addition theorem [(A6), \cite{BKL2}] for the Bessel functions (see also \cite{Wats.6})
\begin{equation}
    j_m\big(k_\perp(x_++a_+),k_\perp(x_-+a_-)\big)=\sum_{n=-\infty}^\infty j_{m-n}(k_\perp a_+,k_\perp a_-)j_n(k_\perp x_+,k_\perp x_-).
\end{equation}
Using this formula, we deduce
\begin{equation}
    \La_{\al\be}(\mathbf{a})\equiv \La(s,m,k_3,k_\perp;s',m',k'_3,k'_\perp;\mathbf{a})=e^{ik_3a_3}j_{m'-m}(k_\perp a_+,k_\perp a_-) \de_{ss'} \frac{2\pi}{L_z}\de(k_3-k'_3) \frac{\pi}{R}\de(k_\perp-k'_\perp).
\end{equation}
Inasmuch as
\begin{equation}
    \hat{V}(\mathbf{a}) \hat{S}_{T/2,-T/2}[j^\mu(t,\spx)] \hat{V}^{-1}(\mathbf{a})= \hat{S}_{T/2,-T/2}[j^\mu(t,\spx-\mathbf{a})],
\end{equation}
the amplitude of creation of a twisted photon by the current $j^\mu(t,\spx-\mathbf{a})$ is given by (see [(30), \cite{BKL2}])
\begin{equation}\label{ampl_transl}
    A(a;s,m,k_3,k_\perp)=\sum_{n=-\infty}^\infty e^{-ik_3a_3}j^*_{m-n}(k_\perp a_+,k_\perp a_-)A(0;s,n,k_3,k_\perp).
\end{equation}
Below, we shall use this formula to derive the probability of radiation of twisted photons by a bunch of identical charged particles moving along parallel trajectories.

\paragraph{Radiation by a bunch of particles.}

Let the distribution of identical particles in the bunch be axially symmetric with respect to the detector axis at some instant of time and described by the function $f(r/\s)$, where $\s$ characterizes the width of the bunch. The normalization condition looks as
\begin{equation}
    2\pi\int_0^\infty dr rf(r/\s)=N,
\end{equation}
where $N$ is the number of particles. Suppose that the particle trajectories pass one into another by shifts by the vectors of the form $\mathbf{a}=(a_1,a_2,0)$, i.e., $a_3=0$.

Assuming the coherent addition of radiation amplitudes, we find the total radiation amplitude
\begin{equation}\label{coh_add}
    \int da_1 da_2 f(|a_+|/\s) A(a;s,m,k_3,k_\perp)=2\pi\int_0^\infty drr f(r/\s) J_0(k_\perp r) A(0;s,m,k_3,k_\perp).
\end{equation}
In integrating in \eqref{coh_add} over the azimuth angle, only one term at $n=m$ in the sum \eqref{ampl_transl} survives. The probability of coherent radiation by such a bunch of particles becomes
\begin{equation}\label{prob_coh}
\begin{split}
    dP^c_f(s,m,k_3,k_\perp)=&\,N^2\vf^2(k_\perp\s)dP_1(s,m,k_3,k_\perp),\\
    \vf(k_\perp\s):=&\,\frac{2\pi}{N}\int_0^\infty drr f(r/\s) J_0(k_\perp r),
\end{split}
\end{equation}
where $dP_1(s,m,k_3,k_\perp)$ is the probability of radiation of twisted photons by one particle moving along the trajectory with the parameter $\mathbf{a}=0$. The interference factor obeys the normalization condition $\vf(0)=1$. The average projection of the total angular momentum onto the detector axis and the projection of the total angular momentum per photon take the form
\begin{equation}\label{ell_coh}
\begin{split}
    dJ_{3f}^c(s,k_3,k_\perp)&=\sum_{m=-\infty}^\infty mdP^c_f(s,m,k_3,k_\perp)=N^2\vf^2(k_\perp\s) dJ_{31}(s,k_3,k_\perp),\\ \ell^c_f(s,k_3,k_\perp)&=dJ_{3f}^c(s,k_3,k_\perp)/dP^c_f(s,k_3,k_\perp)=\ell_1(s,k_3,k_\perp),
\end{split}
\end{equation}
where $dJ_{31}(s,k_3,k_\perp)$ and $\ell_1(s,k_3,k_\perp)$ are the respective quantities for the radiation produced by one particle.

The coherent addition of radiation amplitudes of different particles occurs when the wavelength of radiated photons is larger or of the order of the size of a bunch of particles. Besides, the formulas for coherent radiation can be used in semiclassical description of radiation created by a wave packet of a charged particle. Some properties of coherent radiation of twisted photons such as its infrared asymptotics and the selection rules for symmetrical sources were investigated in \cite{BKL3}. The studies of scattering processes of particles with the wave functions localized in space and possessing nontrivial phases can be found, for example, in \cite{KotSerSchi,BBTq1,BBTq2,BBTZ,KarlJHEP,KKSS,KarlJHEPwp}.

As far as incoherent radiation is concerned, the radiation probabilities are summed up, rather than the amplitudes. Therefore,
\begin{multline}\label{incoh_add}
    dP^{nc}_f(s,m,k_3,k_\perp)=\int da_1 da_2 f(-a) |A(a;s,m,k_3,k_\perp)|^2=\sum_{n,n'=-\infty}^\infty \int da_1 da_2 f(|a_+|/\s)\times\\
    \times A(0;s,n,k_3,k_\perp)A^*(0;s,n',k_3,k_\perp) j^*_{m-n}(k_\perp a_+,k_\perp a_-) j_{m-n'}(k_\perp a_+,k_\perp a_-).
\end{multline}
Notice that, in the case of incoherent radiation, the parameter $a_3$ can be nonzero for different trajectories. The following sum rule is fulfilled
\begin{equation}\label{sum_rule}
    \sum_{m=-\infty}^\infty dP^{nc}_f(s,m,k_3,k_\perp)=N\sum_{m=-\infty}^\infty dP_1(s,m,k_3,k_\perp),
\end{equation}
which is valid for any bunch of particles moving along parallel trajectories. This property stems from the addition theorem [(A6), \cite{BKL2}]
\begin{equation}
    \sum_{m=-\infty}^\infty j^*_{m-n}(k_\perp a_+,k_\perp a_-) j_{m-n'}(k_\perp a_+,k_\perp a_-)=j_{n-n'}(0,0)=\de_{n,n'}.
\end{equation}
A less formal explanation of this property is that, having summed over $m$, the probability of photon radiation becomes invariant with respect to translations of the current $j^{\mu}(x)$. Then the radiation probabilities summed are the same for all the particles in the bunch and simply add up in the case of incoherent radiation.

Taking into account the relation
\begin{equation}
    \sum_{m=-\infty}^\infty mj^*_{m-n}(k_\perp a_+,k_\perp a_-) j_{m-n'}(k_\perp a_+,k_\perp a_-)=n\de_{nn'}+\frac12(k_\perp a_+\de_{n,n'+1} +k_\perp a_-\de_{n,n'-1}),
\end{equation}
we obtain
\begin{equation}
\begin{split}
    dJ^{nc}_{3f}(s,k_3,k_\perp)=&\,N dJ^{nc}_{31}(s,k_3,k_\perp) \\
    &+\re\sum_{n=-\infty}^\infty A(0;s,n,k_3,k_\perp)A^*(0;s,n-1,k_3,k_\perp) \int da_1 da_2 f(-a) k_\perp a_+.
\end{split}
\end{equation}
If the bunch dipole moment defined with respect to the center of the bunch is zero, which is valid, for example, for an axially symmetric bunch, then the last term vanishes. In that case, we have the another sum rule
\begin{equation}\label{J3_incoh}
    dJ^{nc}_{3f}(s,k_3,k_\perp)=N dJ_{31}(s,k_3,k_\perp),\qquad\ell^{nc}_f(s,k_3,k_\perp)=\ell_1(s,k_3,k_\perp).
\end{equation}
The relations \eqref{ell_coh}, \eqref{J3_incoh} indicate that the angular momentum per photon is a rather robust characteristic of radiation produced by axially symmetric bunches.

If the bunch is axially symmetric, the integral over the azimuth angle in \eqref{incoh_add} is readily evaluated
\begin{equation}
    \int_0^{2\pi} d\vf j^*_{m-n}(k_\perp a_+,k_\perp a_-) j_{m-n'}(k_\perp a_+,k_\perp a_-)=2\pi\de_{n,n'} J^2_{m-n}(k_\perp|a_+|),
\end{equation}
where $\vf=\arg a_+$. Then the radiation probability of twisted photons created by such a bunch of particles takes the form
\begin{equation}\label{incoh_red_gen}
\begin{split}
    dP^{nc}_f(s,m,k_3,k_\perp)=&\,N\sum_{n=-\infty}^\infty f_{m-n}(k_\perp\s) dP_1(s,n,k_3,k_\perp),\\
    f_m(k_\perp\s):=&\,\frac{2\pi}{N}\int_0^\infty dr rf(r/\s) J^2_m(k_\perp r).
\end{split}
\end{equation}
It is clear that $f_m(0)=\de_{m,0}$ and $f_m(x)=f_{-m}(x)$. Moreover, for $x\gg\max(1,|m|)$, we have the asymptotics
\begin{equation}
    f_{m}(x)\simeq\frac{1}{\pi x}\frac{2\pi\s^2}{N}\int_0^\infty dr f(r).
\end{equation}
This asymptotics does not depend on $m$. The sum rule \eqref{sum_rule} leads to the normalization condition for the interference factor
\begin{equation}
    \sum_{m=-\infty}^\infty f_m(x)=1.
\end{equation}
Using these properties, one can check once again the validity of general formulas \eqref{J3_incoh} for axially symmetric bunches.

If $dP_1(m)\sim\de_{m,m_0}$, which holds, for example, for a charged particle moving along an ideal helix \cite{Rubic17,KatohPRL,TaHaKa,SasMcNu,BKL2}, then
\begin{equation}
    dP^{nc}_f(s,m,k_3,k_\perp)=N f_{m-m_0}(k_\perp\s) dP_1(s,m_0,k_3,k_\perp).
\end{equation}
In the dipole approximation, the forward radiation created by one particle is mainly concentrated at $m=\pm1$ \cite{BKL2}. In that case,
\begin{equation}\label{dipole_forw}
    dP^{nc}_f(s,m,k_3,k_\perp)\approx N f_{m-1}(k_\perp\s) dP_1(s,1,k_3,k_\perp)+ N f_{m+1}(k_\perp\s) dP_1(s,-1,k_3,k_\perp).
\end{equation}
As we see, the finite width of a particle bunch results in spreading of the incoherent radiation distribution over $m$ in comparison with the radiation probability of twisted photons produced by one point particle.

If $f(r/\s)$ is small for $r\gtrsim\s$, one can neglect the spreading of distribution over $m$ when
\begin{equation}\label{nonspread_cond}
    k_\perp\s\lesssim1.
\end{equation}
It was shown in \cite{BKL2,BKL3} that, in the ultrarelativistic case, the major part of twisted photons is radiated with
\begin{equation}
    n_\perp:=k_\perp/k_0\approx \vk/\gamma,
\end{equation}
where
\begin{equation}
    \vk=\max(1,K),\qquad K=\beta_\perp\gamma,
\end{equation}
and $K$ is the undulator strength parameter, $\beta_\perp>0$ is a characteristic value of the velocity component perpendicular to the detector axis, and $\ga$ is the Lorentz factor of a radiating particle. Then the spreading of distribution over $m$ caused by a finite transverse size of the bunch can be neglected provided that
\begin{equation}\label{1part_cond}
    k_0\s \vk\lesssim\gamma.
\end{equation}
Roughly, this estimate says that one can use the one-particle radiation probability distribution to describe the radiation of twisted photons when the wavelength of radiated photons is larger than the transverse size of a bunch, divided by the Lorentz factor. For example, for the forward radiation produced by charged particles in undulators, we have (see, e.g., [(117), \cite{BKL2}])
\begin{equation}
    \frac{2n\omega\ga\s \vk}{1+K^2+ \vk^2}\lesssim1,
\end{equation}
where $n$ is a harmonic number and $\omega$ is a circular oscillation frequency of particles in the undulator. In Sec. \ref{Und_Rad}, we shall consider the radiation by particle bunches in undulators in more detail.

\paragraph{Interference factors.}

\begin{figure}[!t]
\centering
\includegraphics*[width=0.95\linewidth]{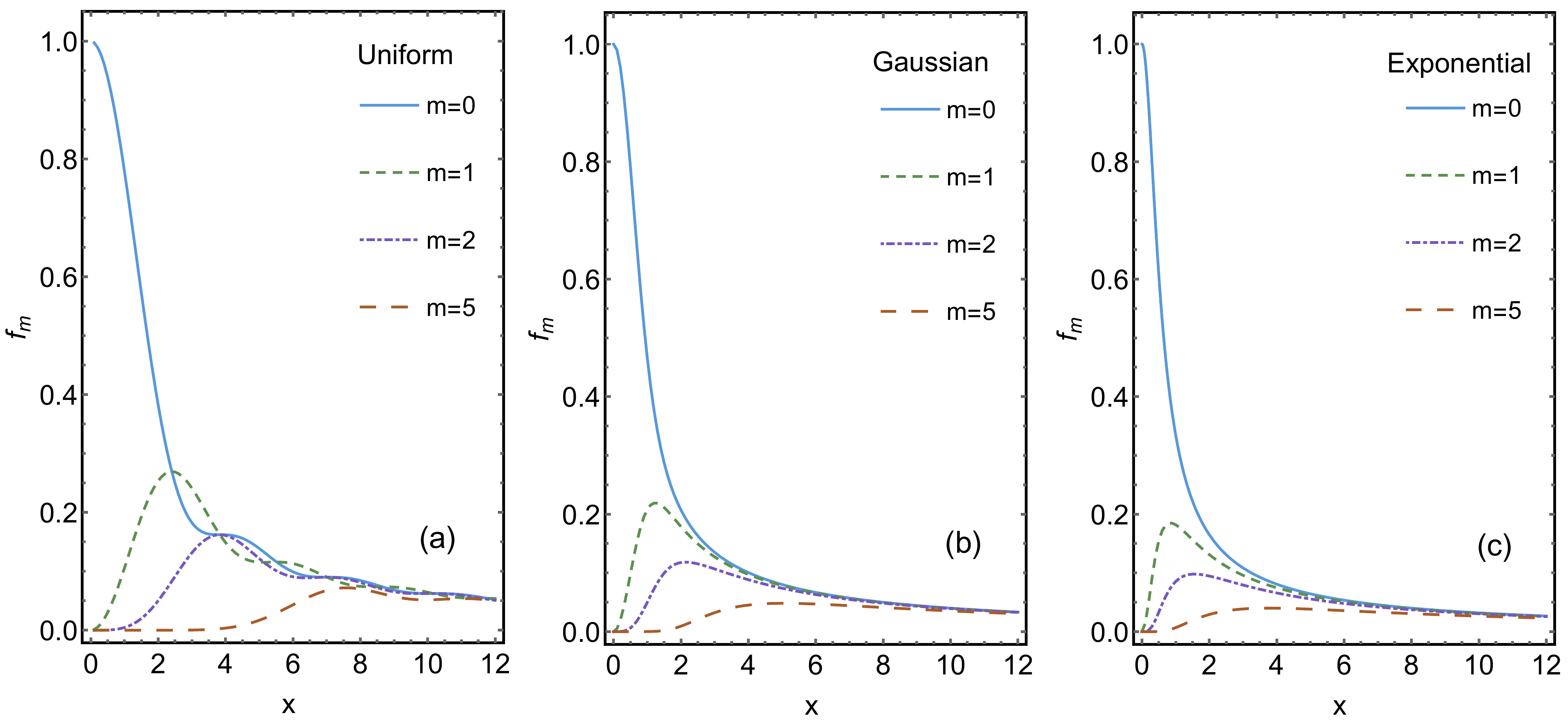}
\caption{{\footnotesize The interference factor $f_m(x)$ for the different bunch profiles.}}
\label{inter_factor_plots}
\end{figure}

Let us find the explicit expressions for the interference factors for simple profile functions $f(r/\s)$.

a) Uniform distribution
\begin{equation}
    f(r/\s)=\frac{N}{\pi \s^2}\theta(\s-r).
\end{equation}
Then
\begin{equation}
    \vf(x)=2J_1(x)/x.
\end{equation}
This expression vanishes when
\begin{equation}\label{zero_rad}
    x=k_\perp\s=\al_1^{(n)},
\end{equation}
where $\al_1^{(n)}$ are the zeros of the Bessel function $J_1(x)$, i.e., the coherent bunch of charged particles does not radiate the twisted photons with $k_\perp$ satisfying \eqref{zero_rad}. For $x\gtrsim 3$, the corresponding factor in the probability of coherent radiation \eqref{prob_coh} drops as $1/x^{3}$.

As for incoherent radiation, the interference factor is written as
\begin{equation}
    f_m(x)=J^2_{|m|}(x) -2\frac{|m|}{x} J_{|m|}(x) J_{|m|+1}(x)  + J^2_{|m|+1}(x).
\end{equation}
This expression drops as $2/(\pi x)$ as $x$ tends to infinity.

b) Gaussian bunch
\begin{equation}
    f(r/\s)=\frac{N}{2\pi \s^2} e^{-r^2/(2\s^2)}.
\end{equation}
In this case, the coherent interference factor becomes
\begin{equation}\label{int_fac_coh_g}
    \vf(x)=e^{-x^2/2}.
\end{equation}
This expression is a monotonically decreasing function of $x=k_\perp\s$ and nonvanishing. For $x\gtrsim3$, the coherent radiation \eqref{prob_coh} is virtually absent.

The incoherent interference factor takes the form
\begin{equation}
    f_m(x)=e^{-x^2} I_{|m|}(x^2),
\end{equation}
where $I_\nu(z)$ is the modified Bessel function of the first kind. For $x$ large, the interference factor decreases as $1/(\sqrt{2\pi}x)$.

c) Exponential profile (see, e.g., \cite{KarlJHEPwp})
\begin{equation}
    f(r/\s)=\frac{N}{2\pi\s^2}e^{-r/\s}.
\end{equation}
The interference factor for coherent radiation is
\begin{equation}
    \vf(x)=(1+x^2)^{-3/2}.
\end{equation}
For $x\gtrsim3$, the corresponding factor entering into \eqref{prob_coh} declines as $1/x^6$. The interference factor for incoherent radiation reads as
\begin{equation}
    f_m(x)= 2\frac{\Ga(|m|+3/2)}{\pi^{1/2}|m|!}
    x^{2|m|} F(|m|+1/2,|m|+3/2;2|m|+1;-4 x^2).
\end{equation}
For $x$ large, this quantity drops as $1/(\pi x)$. The plots of the functions $f_m(x)$ are presented in Fig. \ref{inter_factor_plots}.

\paragraph{Radiation at a small angle.}

\begin{figure}[!t]
\centering
\includegraphics*[width=0.95\linewidth]{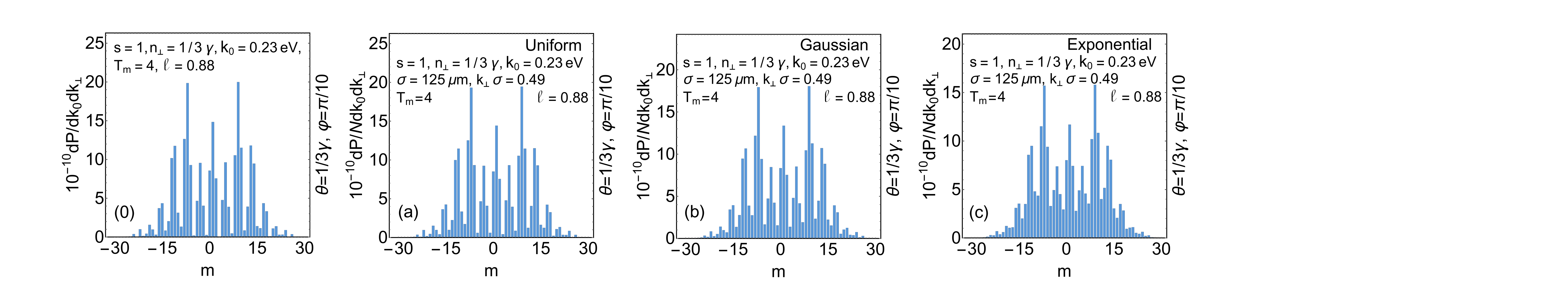}\\
\includegraphics*[width=0.95\linewidth]{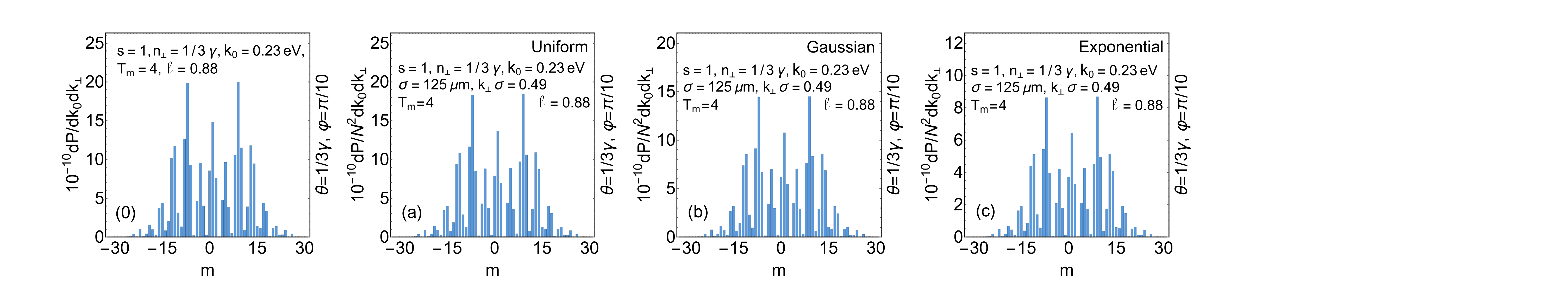}
\caption{{\footnotesize The first line: $0)$ The radiation probability of twisted photons at the first harmonic produced by one electron in the planar undulator with the parameters: the electron Lorentz factor $\ga=100$, the undulator period $\la_0=10$ cm, the number of undulator sections $N_s=40$, the strength of the magnetic field $H=100$ G, the undulator strength parameter $K=6.6\times10^{-2}$. The photons are radiated in the mid-wavelength infrared spectral range. In accordance with \cite{BKL2}, the distribution over $m$ reveals a periodic structure with the period $T_m=4$. a) The radiation probability per electron by an axially symmetric incoherent bunch with the uniform radial distribution of particles. The periodic structure is clearly seen. b) The same but for the Gaussian radial distribution of particles in the bunch. c) The same but for the exponential radial distribution of particles in the bunch. The second line: The same as in the first line but for coherent addition of radiation amplitudes. The form of the distribution $dP(m)$ is almost the same for all the bunch profiles and coincides with the one-particle distribution up to a common factor.}}
\label{small_ang_plots}
\end{figure}

The above formulas can be easily generalized to the case when the detector axis is directed at a small angle $\theta$ to the axis of an axially symmetric bunch. In that case, choosing properly the system of coordinates, the trajectories of particles in the bunch are obtained from one trajectory by means of the translations of the form
\begin{equation}\label{transl_adv}
    a_+=\tau_1\cos\theta+i\tau_2,\qquad a_3=\tau_1\sin\theta,
\end{equation}
where $\tau_{1,2}\in \mathbb{R}$ are the transformation parameters.

For incoherent radiation, the probability to detect the twisted photon is given by
\begin{multline}\label{incoh_add_adv}
    dP^{nc}_{f}(s,m,k_3,k_\perp)=\int d\tau_1 d\tau_2 f(|\tau_+|/\s) |A(a(\tau);s,m,k_3,k_\perp)|^2=\sum_{n,n'=-\infty}^\infty \int d\tau_1 d\tau_2 f(|\tau_+|/\s)\times\\
    \times A(0;s,n,k_3,k_\perp)A^*(0;s,n',k_3,k_\perp) j^*_{m-n}\big(k_\perp a_+(\tau),k_\perp a_-(\tau)\big) j_{m-n'}\big(k_\perp a_+(\tau),k_\perp a_-(\tau)\big).
\end{multline}
If
\begin{equation}\label{small_angl}
    k_\perp\s\theta^2/2\ll1,\qquad \theta^2\ll1,
\end{equation}
then, using the expansions,
\begin{equation}
\begin{split}
    a_\pm&=\tau_\pm-\tau_1\theta^2/2+\cdots,\\
    j_k(k_\perp a_+,k_\perp a_-)&=j_k(k_\perp \tau_+,k_\perp \tau_-)+\frac{k_\perp\tau_1\theta^2}{4}\big[j_{k+1}(k_\perp \tau_+,k_\perp \tau_-) -j_{k-1}(k_\perp \tau_+,k_\perp \tau_-) \big]+\cdots,
\end{split}
\end{equation}
we deduce that one can set $\cos\theta=1$ in the expression \eqref{transl_adv} for $a_\pm$, provided $f(r/\sigma)$ is small for $r\gtrsim\s$. As a result, the integral \eqref{incoh_add_adv} is reduced to \eqref{incoh_add}, and we revert to formula \eqref{incoh_red_gen}. Thus, in the case of incoherent radiation of twisted photons by axially symmetric bunches with the symmetry axis tilted from the detector axis by a small angle, one can still use formula \eqref{incoh_red_gen}.

The case of coherent radiation is a little more sophisticated. The radiation amplitude is written as
\begin{equation}\label{coh_add_adv}
\begin{split}
    \int d\tau_1 d\tau_2 f(|\tau_+|/\s) A(a(\tau);s,m,k_3,k_\perp)=& \int_0^\infty drr f(r/\s)\sum_{n=-\infty}^\infty A(0;s,n,k_3,k_\perp)\times\\
    &\times\int_0^{2\pi}d\vf e^{-i[k_3 r\sin\theta\cos\vf+(m-n)\arg a_+(\tau)]} J_{m-n}\big(k_\perp |a_+(\tau)|\big),
\end{split}
\end{equation}
where $r:=|\tau_+|$ and $\vf$ is the azimuth angle in the plane $(\tau_1,\tau_2)$. Under the assumptions \eqref{small_angl}, we can take $a_+=\tau_+$ in the expression \eqref{coh_add_adv}. Then the integral over $\vf$ is the Bessel function, and we have
\begin{equation}
    \int d\tau_1 d\tau_2 f(|\tau_+|/\s) A(a(\tau);s,m,k_3,k_\perp)=N\sum_{n=-\infty}^\infty \vf_{m-n}(k_\perp\s,k_3\s\sin\theta) A(0;s,n,k_3,k_\perp),
\end{equation}
where the interference factor takes the form
\begin{equation}
    \vf_{m}(k_\perp\s,k_3\s\sin\theta):=i^{-m}\frac{2\pi}{N}\int_0^\infty drrf(r/\s)J_{m}(k_3r\sin\theta)J_{m}(k_\perp r).
\end{equation}
Obviously,
\begin{equation}
    \vf_m(x,y)=\vf_m(y,x),\qquad \vf_m(x,y)=(-1)^m\vf_{-m}(x,y),\qquad \vf_m(0,0)=\de_{m,0}.
\end{equation}
The probability of coherent radiation of twisted photons becomes
\begin{equation}\label{prob_coh_adv}
    dP^c_f(s,m,k_3,k_\perp)=N^2\Big|\sum_{n=-\infty}^\infty \vf_{m-n}(k_\perp\s,k_3\s\sin\theta)A(0;s,n,k_3,k_\perp)\Big|^2.
\end{equation}
If $f(r/\s)$ decreases rapidly for $r\gtrsim\s$, and
\begin{equation}\label{y_sml}
    k_3\s\theta\ll1,
\end{equation}
then, up to contributions of the order  $f(10)$, we come back to formulas \eqref{coh_add}, \eqref{prob_coh}.

In the dipole approximation, the radiation amplitude of a twisted photon differs from zero only for $m=\pm1$. In that case, only the two terms with $n=\pm1$ are left in the sum \eqref{prob_coh_adv}. When the particles move along an ideal helix with the axis coinciding with the detector axis, the only one term survives in the sum \eqref{prob_coh_adv}. It corresponds to $|n|=n_0$, where $n_0$ is the harmonic number, and the sign of $n$ is determined by the helix chirality. The examples of radiation probability distributions of twisted photons recorder at a small angle and created by axially symmetric particle bunches are presented in Fig. \ref{small_ang_plots}.

\begin{figure}[!t]
\centering
\includegraphics*[width=0.95\linewidth]{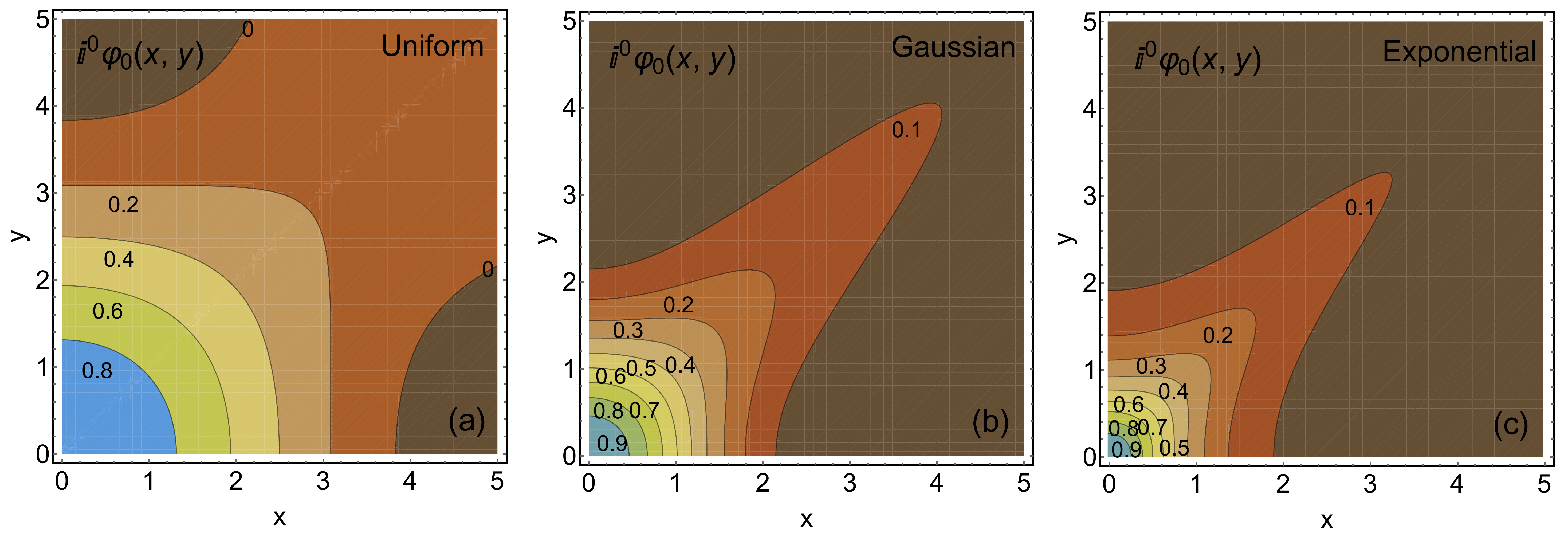}\\
\includegraphics*[width=0.95\linewidth]{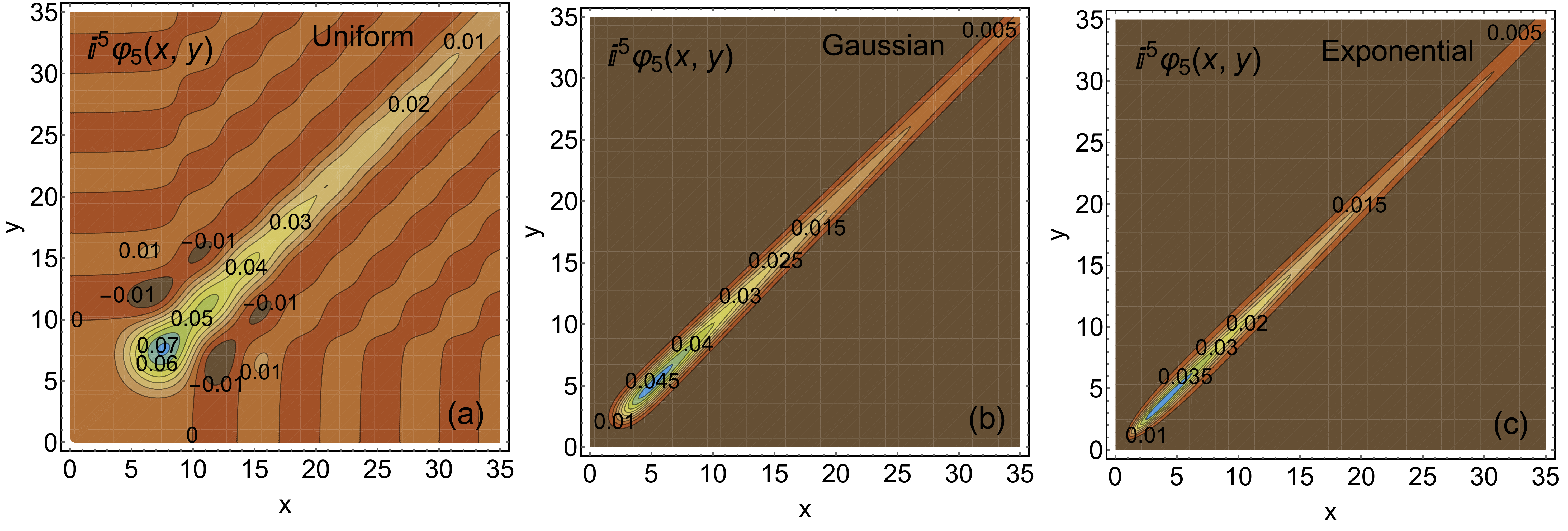}
\caption{{\footnotesize The interference factor $i^m\vf_m(x,y)$ for the different bunch profiles. The first line: $m=0$. The second line: $m=5$.}}
\label{inter_factor_ang_plots}
\end{figure}

Let us present the explicit expressions for $\vf_{m}(x,y)$ for the bunch profiles considered above \cite{PruBryMar2}:
\begin{equation}\label{int_fac_coh_angl}
\begin{split}
    a)&\;\vf_{m}(x,y)=2i^{-m}\frac{xJ_{|m|}(y)J_{|m|-1}(x)-yJ_{|m|}(x)J_{|m|-1}(y)}{y^2-x^2},\\
    b)&\;\vf_{m}(x,y)=i^{-m}e^{-(x^2+y^2)/2}I_{|m|}(xy),\\
    c)&\;\vf_{m}(x,y)=-\frac{2}{\pi}i^{-m}\frac{Q^1_{|m|-1/2}\big((1+x^2+y^2)/(2xy)\big)}{\sqrt{xy(1+(x+y)^2)(1+(x-y)^2)}},
\end{split}
\end{equation}
where $x:=k_\perp\s$, $y:=k_3\s\sin\theta$, and the associated Legendre function of the second kind $Q^1_{|m|-1/2}(x)$ is defined in such a way that it is real and analytic if $x>1$. In the cases (b) and (c), the interference factors decline rapidly to zero out of the diagonal $x=y$, i.e., for
\begin{equation}
    k_\perp/k_3\neq\theta,
\end{equation}
provided $m\neq0$. This implies that if $x\neq y$, then only the term with $n=m$ survives in the sum \eqref{prob_coh_adv}, and the spreading of radiation probability distribution over $m$ is negligible. In the case (a), the modulus of $\vf_m(x,y)$ also reaches the maximal value at $x=y$, but its decrease out of the diagonal is not so drastic. For $m=0$, besides the diagonal, the expressions \eqref{int_fac_coh_angl} are not small in the region $x\lesssim1$, $y\lesssim1$. Even on the diagonal the quantity $|\vf_0(x,x)|\gg|\vf_m(x,x)|$, where $m\neq0$ and $x\lesssim1$, and so the spreading of the corresponding one-particle radiation probability distribution is negligible in this case too (see Fig. \ref{small_ang_plots}).

Notice that the expression (b) turns into \eqref{int_fac_coh_g} when not only $y\ll1$, but $xy\ll1$. At first glance, this contradicts the statement made below formula \eqref{y_sml}. However, this controversy is resolved if one observes that, in the case at hand, the exact expression (b) is very small for $x$ large. The error arising in passing from \eqref{prob_coh_adv} to \eqref{prob_coh} is negligibly small. The plots of functions $\vf_{m}(x,y)$ are given in Fig. \ref{inter_factor_ang_plots}.

\paragraph{Transition to a continuous flow.}

\begin{figure}[!t]
\centering
\includegraphics*[width=0.95\linewidth]{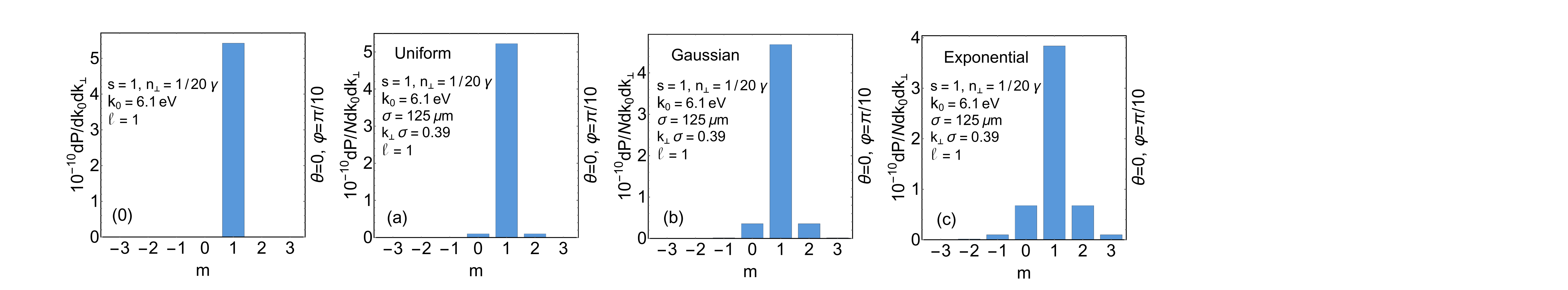}\\
\includegraphics*[width=0.95\linewidth]{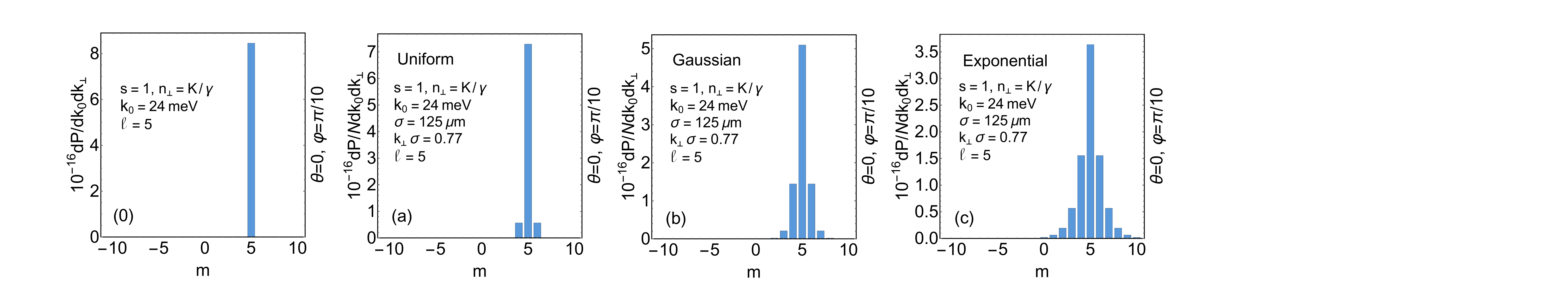}\\
\includegraphics*[width=0.95\linewidth]{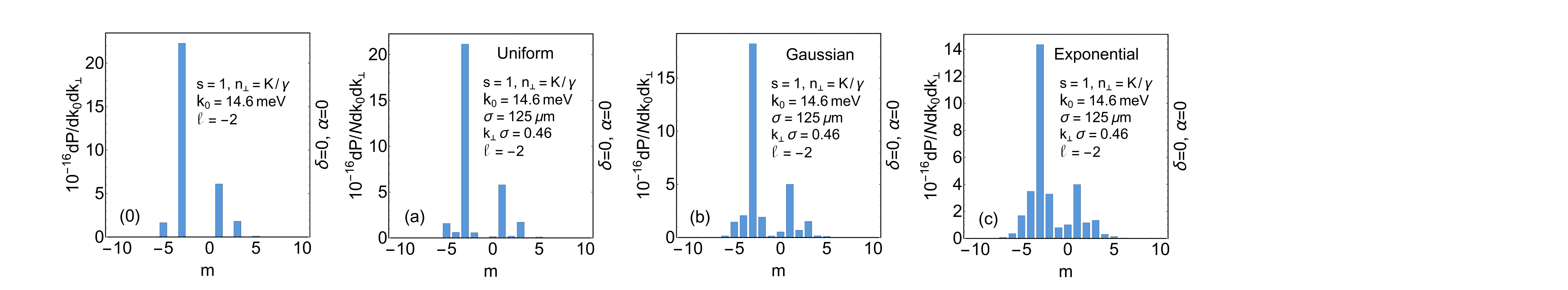}
\caption{{\footnotesize The first line: $0)$ The probability of forward radiation of a twisted photon at the first harmonic by one electron in the planar undulator with the parameters: the electron Lorentz factor $\ga=500$, the undulator period $\la_0=10$ cm, the number of undulator sections $N_s=40$, the strength of the magnetic field $H=100$ G, the undulator strength parameter $K=6.6\times10^{-2}$. The photons are radiated in the middle ultraviolet spectral range. a) The forward radiation probability per electron by an axially symmetric incoherent bunch with the uniform radial distribution of particles. b) The same but for the Gaussian radial distribution of particles in the bunch. c) The same but for the exponential radial distribution of particles in the bunch. The second line: The same as in the first line but for an ideal helical wiggler with $\ga=100$, $\la_0=10$ cm, $N_s=40$, $K=5$. The fifth harmonic is considered. The photons are radiated in the THz spectral range. The forward radiation produced by one particle obeys the selection rule $m=n$, where $n$ is the harmonic number. The finite width of a particle bunch results in spreading of the distribution over $m$. The third line: The same as in the second line but for the planar wiggler with the parameters $\ga=100$, $\la_0=10$ cm, $N_s=40$, $K=5$. The third harmonic is considered. The photons are radiated in the THz spectral range. The forward radiation produced by one particle in the planar wiggler obeys the selection rule: $n+m$ is an even number \cite{BKL2}. The finite width of a particle bunch leads to violation of this selection rule. Nevertheless, the peaks at $m=\{-3,1,3\}$ are still pronounced.}}
\label{on_axis2_plots}
\end{figure}

We implicitly assumed above that the bunch of particles can be replaced by a continuous flow, i.e., one can pass from the summation over particles to the integration over a continuous distribution. This transition is justified in the case when the radiation by a particle wave packet or a coherent bunch of quantum particles are considered in the semiclassical approximation. In this approximation, the radiation is such as if it were generated by the current of a charged fluid with the properties (the charge density and the velocity) determined by the wave function. Then the radiation produced is coherent, i.e., one should use the formulas for coherent radiation. Furthermore, the bunch of particles can be replaced by a continuous flow in the case when
\begin{equation}
    k_\perp|\De a_+|\ll1,
\end{equation}
where $|\De a_+|$ is a distance between neighboring particles. This is valid for both coherent and incoherent radiation generated by particle bunches.

\section{Undulator radiation}\label{Und_Rad}

\paragraph{Ordinary undulators.}

As an example of applications of the general formulas obtained above, we consider the radiation of twisted photons by a bunch of charged particles in undulators. The radiation of twisted photons by one charged particle in an undulator was thoroughly investigated in \cite{SasMcNu,HMRR,HeMaRo,AfanMikh,BordKN,KatohPRL,TaHaKa,Rubic17,BKL2}. We will use the notation and the results presented in Sec. 5 of \cite{BKL2} (see also \cite{Bord.1}).

From the physical point of view, the most interesting case is the incoherent addition of radiation amplitudes of separate particles in the bunch. It is this situation which is commonly realized in experiments. The analysis carried out in the previous section indicate that the spreading of distributions of radiated twisted photons over $m$ is small when \eqref{1part_cond} is satisfied. In that case, the one-particle results provide a good approximation for the radiation by particle bunches. In Figs. \ref{small_ang_plots}, \ref{on_axis2_plots}, the distributions of twisted photons over $m$ radiated by an electron bunch with a round cross section $\s=125$ $\mu$m are presented. Such bunches of electrons and positrons are created, for example, in the electron-positron collider VEPP-2000 \cite{RPPPDG2018}. Despite the spreading of distributions, the projection of the angular momentum per photon $\ell$ remains the same as for the one-particle distribution (see \eqref{J3_incoh} and the plots in Figs. \ref{small_ang_plots}, \ref{on_axis2_plots}). The selection rules that hold for the forward radiation produced by one particle in the helical and planar undulators are violated for the radiation generated by a bunch of particles. However, if $k_\perp\s$ is small, their existence for a one-particle distribution can be tracked out from the plots in Fig. \ref{on_axis2_plots}. As is seen from Fig. \ref{small_ang_plots}, the periodicity in $m$ for undulator radiation at an angle \cite{BKL2} is a more robust property.

\paragraph{Crystalline undulators.}

\begin{figure}[!t]
\centering
\includegraphics*[width=0.325\linewidth]{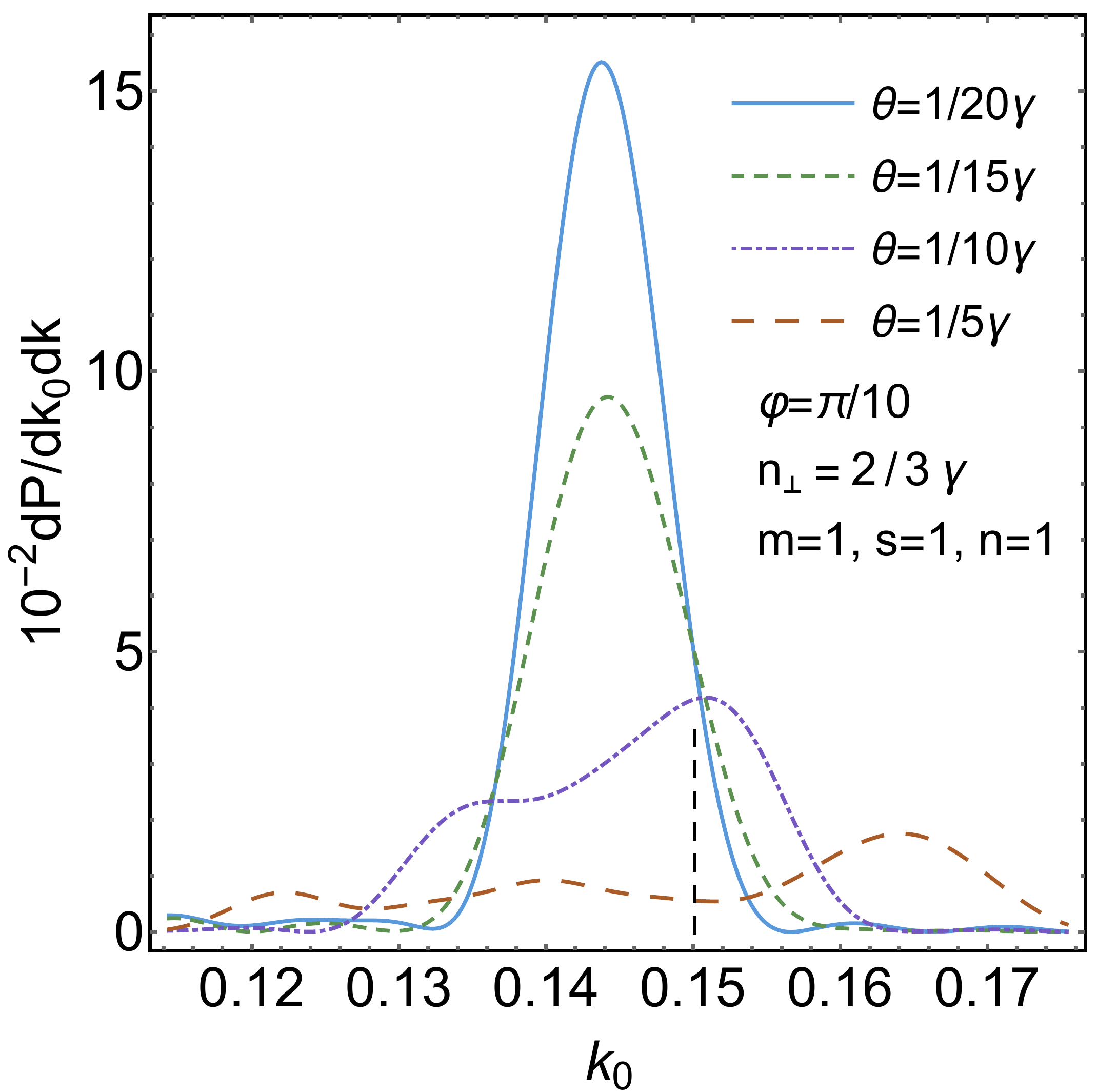}\;\;
\includegraphics*[width=0.5\linewidth]{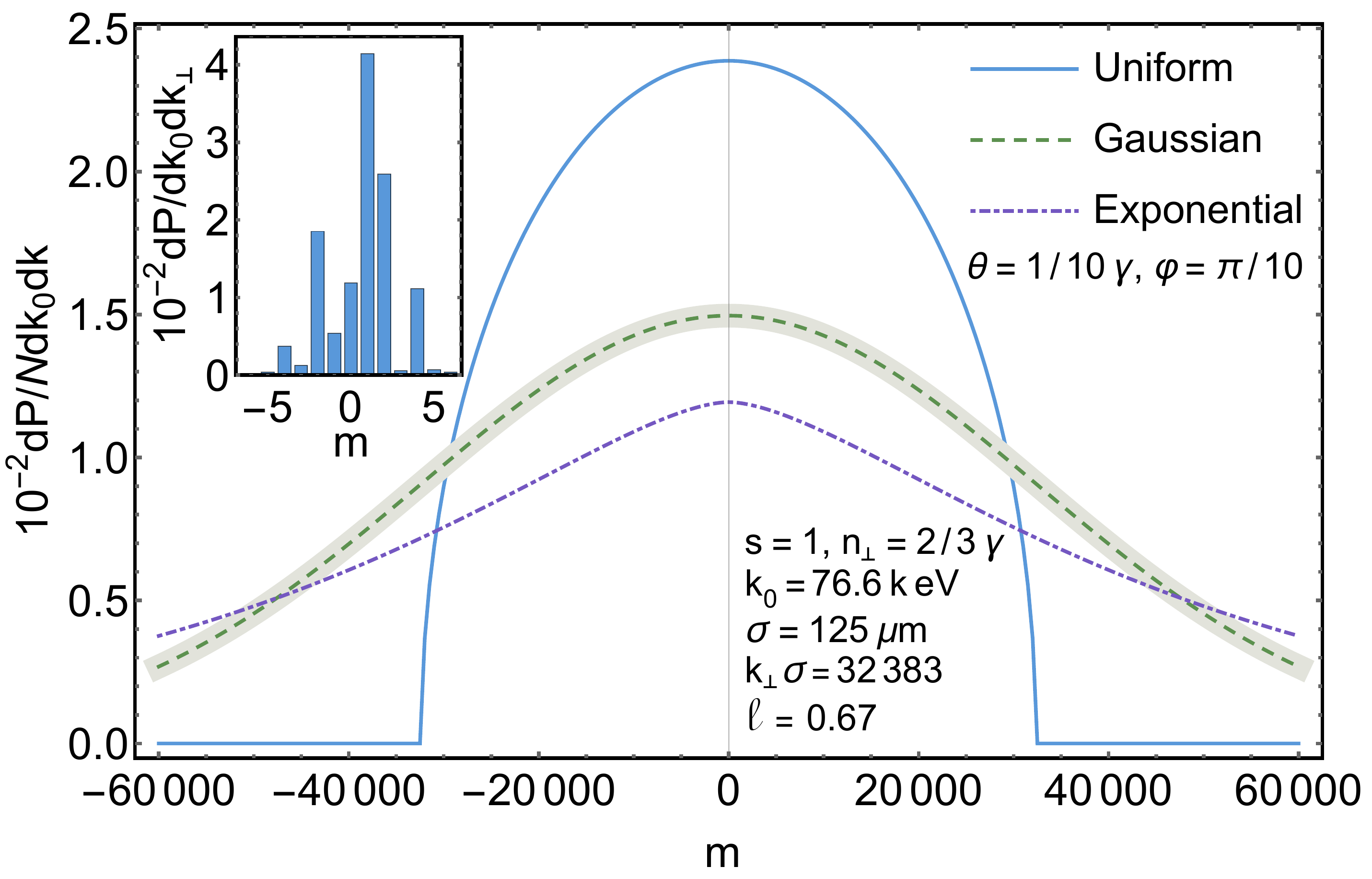}
\caption{{\footnotesize Left panel: The dependence of probability of radiation of twisted photons produced in the crystalline undulator with $K=0.23$ on the photon energy for the different observation angles. The photon energy $k_0=0.15 m=76.6$ keV used on the right panel is depicted. Right panel: The probability per electron of incoherent radiation of twisted photons in the crystalline undulator at $k_0=76.6$ keV (the hard X-rays) for the different axially symmetric bunch profiles. The projection of the angular momentum per photon $\ell$ is the same for all the profiles. The thick line is a curve given by \eqref{dP_large} for the Gaussian bunch. The inset: The one-particle radiation probability distribution of twisted photons over $m$.}}
\label{on_axis2_crys_plots}
\end{figure}

As the second example, we consider the radiation of twisted photons by crystalline undulators \cite{KKSGb}. The trajectory of the electron in such an undulator is given approximately by
\begin{equation}
    x(t)=-a\cos(\omega t),\qquad y(t)=0,\qquad z(t)=\be_\parallel(t-L/2),\qquad t\in[0,L],
\end{equation}
where $\be_\parallel=(1-\ga^{-2})^{1/2}$, $\ga=10^3$, $\omega=2\pi\be_\parallel\la_0^{-1}$, $N_s=15$, $L=2\pi N_s/\omega$, $\la_0=23$ $\mu$m, $a=10d$, $d=0.1$ nm. The parameters are taken from \cite{KKSG}. We suppose that the transverse size of the electron bunch is characterized by $\s=125$ $\mu$m, and the number of particles in the bunch $N=10^{10}$. In that case, the energy of radiated photons is large and $k_\perp\s\gg1$.

If $k_\perp\s\gg1$, then $f_{m}(x)$ depends weakly on $m$ in the region where $f_{m}(x)$ is not exponentially suppressed. Therefore, if the one-particle radiation distribution over $m$ is nonvanishing only in the region $|m|\ll k_\perp\s$, as, for example, in the case of the forward dipole radiation \eqref{dipole_forw}, then \eqref{incoh_red_gen} implies approximately
\begin{equation}\label{dP_large}
    dP^{nc}_f(s,m,k_3,k_\perp)\approx N f_{m}(k_\perp\s) \sum_{n=-\infty}^\infty dP_1(s,n,k_3,k_\perp)= N f_{m}(k_\perp\s) dP_1(s,k_3,k_\perp).
\end{equation}
Thus, in this case, the distribution over $m$ of twisted photons radiated by a bunch of charged particles is universal up to a common factor. The approximation \eqref{dP_large} is rough and does not comply with the exact sum rule \eqref{J3_incoh}. Nevertheless, \eqref{dP_large} reproduces the shape of the curve $dP(m)$ quite well (see Fig. \ref{on_axis2_crys_plots}). One can secure the compliance with the sum rule \eqref{J3_incoh} by making the replacement $f_{m}(k_\perp\s)\rightarrow f_{m-m_0}(k_\perp\s)$ in \eqref{dP_large}, where $m_0$ is found from the requirement
\begin{equation}
    \sum_{m=-\infty}^\infty mf_{m-m_0}(k_\perp\s)=\ell_1(s,k_3,k_\perp),
\end{equation}
and $\ell_1$ is given in \eqref{J3_incoh}. In general, $m_0=m_0(s,k_3,k_\perp)$.

\section{Conclusion}

Let us briefly recapitulate the results. We obtained the general formulas \eqref{prob_coh}, \eqref{incoh_red_gen}, and \eqref{prob_coh_adv} relating the radiation probability of twisted photons produced by axially symmetric bunches to the same quantity for one point particle. The general sum rules \eqref{ell_coh}, \eqref{sum_rule}, and \eqref{J3_incoh} were established. In particular, we found that the projection of the total angular momentum per photon radiated by axially symmetric bunches does not depend on the bunch profile and coincides with the same quantity for the radiation created by one point particle. The relations between the radiation probabilities by particle bunches and the one-particle radiation probability involve the interference factors. We proved some general properties of these interference factors and obtained their explicit form for simple radial bunch profiles. In general, a finiteness of the particle bunch width leads to spreading of the radiation probability of twisted photons over $m$. We found the condition \eqref{nonspread_cond} when this spreading is marginal and the one-particle radiation probability distribution is an adequate approximation for description of radiation by particle bunches.

The general theory was applied to undulator radiation. The two types of undulators were considered: the ordinary undulator and the crystalline one. We showed that the selection rules fulfilled for the forward undulator radiation \cite{Rubic17,KatohPRL,TaHaKa,SasMcNu,BKL2} become violated when the radiation by a bunch of particles is considered. However, when the condition \eqref{nonspread_cond} is satisfied, this violation is negligible. The projection of the total angular momentum per photon is, of course, the same as for the one-particle radiation since it is independent of the axially symmetric bunch profile. It was shown in \cite{BKL2} that the radiation probability distribution over $m$ possesses a periodic structure when the radiation of twisted photons by one particle in undulator is considered at a small angle to the undulator axis. In the present paper, we showed that this property holds for the radiation by particle bunches as well and is less sensitive to the bunch size than the selection rules mentioned above. The plots of the probability of radiation of twisted photons by particle bunches in ordinary undulators are presented in Figs. \ref{small_ang_plots}, \ref{on_axis2_plots}.

As for crystalline undulators, the twisted photons produces by them are much harder than those created in ordinary undulators. Then it follows from the condition \eqref{nonspread_cond} that the one-particle radiation probability distribution over $m$ is subjected to a drastic broadening when a finiteness of a bunch width is taken into account. In that case, the shape of the radiation probability distribution over $m$ becomes universal (see Fig. \ref{on_axis2_crys_plots}), but the projection of the total angular momentum per photon remains the same as for the one-particle radiation.

\paragraph{Acknowledgments.}

We are thankful to D.V. Karlovets and G.Yu. Lazarenko for fruitful conversations. This work is supported by the Russian Science Foundation (project No. 17-72-20013).


\end{document}